# Real-time tuneable bright bonding plasmonic modes in Ga nanostructures.

**RENU RAMAN SAHU,**[1, †] **AND TAPAJYOTI DAS GUPTA**[1, *]

[1]*Laboratory of Advanced Nanostructures for Photonics and Electronics, Department of Instrumentation and Applied Physics, Indian Institute of Science, Bangalore, India-560012*
*\*tapajyoti@iisc.ac.in*

**Abstract:** The precise control of nanogaps is crucial for plasmonic nanoassemblies, where plasmon hybridization is highly sensitive to gap size and geometry. This sensitivity enables fine-tuning of the resonance wavelength and near-field enhancement, offering the potential for advanced optical applications. However, conventional lithographic techniques for gap modulation are constrained to discrete values and face challenges in achieving nanometer order of separations. Such limitations hinder the comprehensive study of plasmon coupling across varying interaction regimes. Overcoming these challenges is essential for advancing nanoplasmonic research and its practical applications. Herein, we demonstrate a tuneable plasmonic device in which real-time tunability of this hybridization mode is achieved via manipulation of the inter-droplet gap of liquid metal nanoparticles by macroscopic physical deformation. In particular, we show that the optical spectra obtained from the sample shift towards higher energy on the application of a linear strain, resulting in an increase of inter-droplet gaps leading to a direct probing of the bright modes in situ. Our method thus offers a novel means of exploring the fundamental concept of real-time tuneable plasmon hybridization as well as tuning of nanoparticle assembly with any desired gap in a controlled manner.

## 1. Introduction

The collective electron oscillations at the metal-dielectric interface, also known as plasmons, are ubiquitously used in applications like, sensing[1–5], spectroscopy[6–10], and energy harvesting[11–15]. In nanoscale metallic nanoparticles, the plasmons are confined to the spatial extent of the plasmonic material, also called localized surface plasmons, and exhibit resonances dependent on the surrounding media, the size, shape, geometry, and orientation of the nanoparticles, thus providing a versatile platform for utilization in nanophotonic devices for control and manipulation of light energy.

Two or more nanoplasmonic systems at spatial proximity lead to the formation of hybrid plasmonic modes with plasmonic responses often different from that of the individual system. The optical scattering processes in an individual plasmonic system are well understood in terms of Mie scattering from nanomaterials with permittivity with negative real parts, the typical characteristic of plasmonic materials. Plasmon hybridization depicts the interaction of the plasmonic modes in the individual systems, leading to a control parameter for the design of plasmonic devices. This principle has allowed for controllable plasmon-enhanced light-matter interactions, applicable in nanolasing[16], gap-tunable Raman spectroscopy[17] gap-plasmon enhanced photoluminescence[18], beam steering, nanofocusing in near-field microscopy[19], and metasurfaces[20]. Therefore, the nanoscale spatial proximity of electron-dense materials is sought after in designing and fabricating plasmonic devices by techniques like dewetting[21,22], e-beam lithography[23], nanostencil-lithography[24], and origami[4,25].

Tuning of plasmonic coupling response is required to obtain a desirable plasmonic response depending on the application the device caters to. The tuning parameters available to obtain a desirable plasmonic response can be optical[26], electronic[27], thermal[28], or geometrical[29]. However, after the fabrication of a plasmonic device, its optical properties are generally fixed. Thus far however, very limited work on real-time plasmonic hybridization tunability is achieved. Dynamical tuning by mechanical deformation has also been achieved with periodic structures of elastomeric substrates. However, as the periodicity changes the optical properties become polarisation-dependent, which limits the usage to a particular orientation of mechanical deformation. On the other hand most of the periodic structures exhibit a red-shift owing to a shift surface lattice resonance[30], further making them dependent on the angle of incidence and also on the lattice parameter. Despite significant progress in both theoretical and experimental studies, advancing this field thus requires the ability to precisely manipulate interparticle distances at the nanometer scale for a randomly distributed nanoparticles and establish direct correlations between these distances and optical spectra, all of which present considerable challenges.

A necessary confirmation of hybridized plasmon mode is the blue shift of optical spectra due to any physical deformation leading to an increase of inter-particle gap. While there have been reports of blue shift of optical spectra, which can be attributed to plasmon hybridization, they employ non-spherical structures like bowtie[24], nanorods[31], bipyramids[32], or cubes[33], which reduces the symmetry of the plasmonic geometry and bring in the polarisation dependence. Ideally, spherical plasmonic morphology with spatial proximity allowing for hybridization and particle distribution with in-plane isotropy will solve the problem of polarisation dependence and allows for polarisation-independent plasmonic spectra.

Gallium as a plasmonic material has been established34 and it is amenable for applications in photonic devices. The bulk plasmon frequency of Ga lies in the deep UV region beyond 13.5 eV34,35, thus making all the lower frequencies accessible for exploration for different applications. Scattering from Ga nanoparticles of radii sub100 nm, according to Mie theory, would have a resonant peak due to dipole surface plasmon (SI Figure 1-4), the lowest energy mode, at around 8.9eV, owing to Frolich's condition (SI Figure 5). Embedding the Ga nanoparticles in a high refractive index medium like PDMS (Poly di methoxysilane, an elastomer) reduces the LSPR peak to around 7 eV which is still in the deep UV region (SI Figure 5). Increasing the sizes of the nanodroplets lowers the surface dipole mode energy to the visible region but at the cost of quality factor (SI Figure 6-7). However, to obtain a plasmonic response in the visible region (1.8eV to 3.1eV) without compromising the quality factor, hybridization of LSP modes is necessitated (SI Figure 8-9) which can be controlled by tuning the gap between Ga nanoparticles. While the fabrication of Ga nanoparticles with uniform dispersity had been a challenge until recently, emerging self-assembly-based techniques overcame this challenge[36.

Herein, we show a novel technique of an elastomeric gap-controlled plasmon hybridized nanostructures of Ga embedded in Polydimethylsiloxane (PDMS) elastomer in a random yet coupled manner. Having a random structure is crucial to mitigate the periodic and angular dependency thus allowing spectral shift purely from plasmon hybridization. The experimental reflectivity spectra exhibit a blue shift with an increasing strain. We show that the blue shift is attributed to the decrease in the plasmon hybridization strength as the inter-droplet gaps increase with mechanical strain. Exploiting the sensitivity of optical scattering to the inner droplet spatial gap through macroscopic spatial deformations, these samples thus have the potential to be used in strain sensors and in soft-tuneable photonics.

## 2. Non-coalescent Ga nanodroplets embedded in PDMS substrate

Thermal evaporation (HHV, Auto 500 thermal evaporator) of liquid Ga (Thermo Scientific Chemicals, 99.999% metal basis, packaged in polyethylene bottle) onto PDMS 10 (PDMS fabrication is detailed in method section) (Figure 1a) results in a structural color due to the formation of Ga nanodroplets (Figure b top view) which gets embedded into the PDMS substrate (Figure 1c cross-sectional SEM image). These droplets, are found to be spherical in shape and in a liquid state, are non-coalescent with sub-nm scale gap and sub-100nm droplet sizes[36].

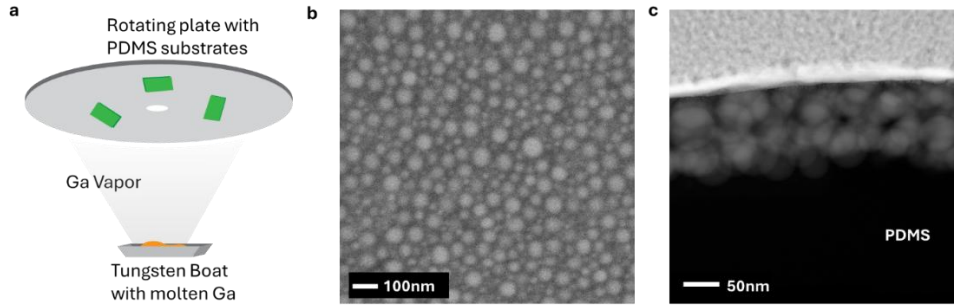

Fig. 1. Fabrication of Ga deposited Polydimethylsiloxane and its morphology. (a) Thermal evaporation of Ga onto PDMS. (b) Top-view SEM image of the sample. (c) Cross-section view obtained from high-angle annular dark field imaging using scanning Transmission Electron Microscope.

As the sample gets linearly strained to about 40.5% of its initial length, the reflectivity spectra obtained from the sample undergo a continuous blue shift with the reflectivity (indicated with arrows in Figure 2a) peak shifting from 590nm to 540nm (Figure 2a-b). On releasing the strain, the reflectivity spectrum red-shifts towards the unstrained spectrum, indicating the reversibility of the spectral change (Figure 2**Error! Reference source not found.**a). Furthermore, a change in the color of the sample (Figure 2b) is observed owing to the shift of spectrum due to mechanical strain.

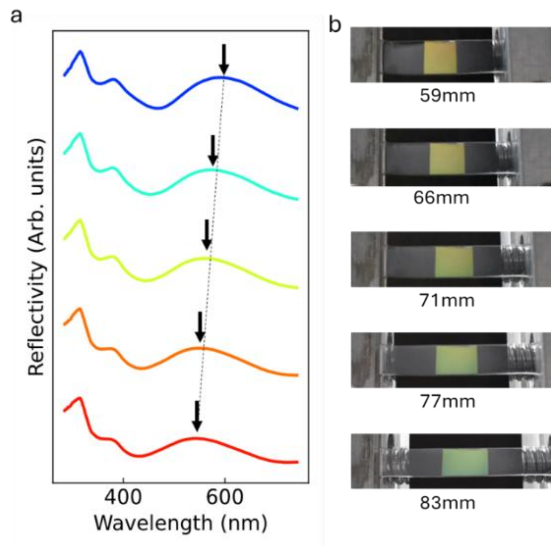

Fig. 2. Reflectivity spectra from the linearly strained sample. (a) Reflectivity spectra of the samples shown in (b) Linearly applied strain along the longitudinal axis of the rectangular sample.

Since the reflectivity spectra and, hence, the color of the sample depends on the structure, it is imperative to analyze the reason for the spectral change from the perspective of structural change. In the following, we propose that the randomly oriented in-plane Ga droplets with almost a monodispersed distribution are equivalent to randomly oriented Ga homodimers with their axes on the substrate plane. Next, we show that to understand the spectral behavior of this system, it is sufficient to study a single dimer mode.

### 3. Ga on PDMS as randomly oriented dimers

In the experimental scenario, the Ga nanodroplets are randomly distributed on the surface of the substrate, on which each adjacent Ga nanodroplet can be thought of as a dimer. Alternatively, the structure is equivalent to dimers on the surface of the substrate with dimer-axes randomly oriented on the substrate surface. Since the light is normally incident, the electric field is parallel to the surface of the substrate, which implies that there would be components of the electric field both parallel and perpendicular to the dimer axis. Therefore, an insight into the single dimer is crucial for understanding the spectral shifts observed experimentally.

### 4. Evolution of Dimer modes

For an analytical treatment of a dimer system[37], one considers the z-axis as the line joining the centers of the individual spheres. The surface charge densities of the individual spheres can be expressed as a linear combination of spherical harmonics $(l, m)$. Due to axial symmetry the modes with different $m$ decouple. For $m = 0$, the electric field is in the direction of the z-axis, parallel to the gap axis. For $m = 1$, the electric field is perpendicular to the gap axis (Figure 3).

When illuminated with plane waves, the charge distribution of the system gets re-oriented according to the plasmonic modes supported by the system. Only the modes with a non-zero dipole moment couple efficiently with plane polarised light and are known as bright -modes. This is why the lowest mode energies correspond to the dipole mode. The other modes characterized by charge distribution with negligible dipole moment are known as dark modes.

Note that the mode energies of the single Ga nanosphere of radius30nm are 3.8eV and 5.7eV, corresponding to the dipole and quadrupole modes, both in the UV region (Figure 3). In the case of a Ga dimer, the energies of the spectral peak change because of plasmon hybridization.

### 5. Bonding and Anti-Bonding modes in Ga dimers

In a dimer the charge distributions of the individual spheres interact with each other. It has been shown that this interaction is a result of coupling between different modes present in the individual Ga spheres. The absorption cross-section spectra is correspondingly modified and have the signatures of the modified plasmon energies. We determine the plasmon modes of the dimers numerically. Although Ga is a UV plasmonic material, the static spatial arrangement of Gallium nanodroplets with sub-10nm inter-droplet spacing results in the hybridisation of local surface plasmon modes. Due to this plasmon hybridization, an optical response in the visible region (1.7eV to 3.1 eV) is observed.

The dimer axis is the line joining the two centers of the Ga spheres. Depending on the direction of electric field polarisation with respect to the dimer axis, the dimer modes are excited, as shown in Figure 3. The bonding modes are those with energy less than that of the single particle, whereas the energy of the anti-bonding mode is higher than that of the single particle. When compared to single particle modes, the lower energy modes in the dimer are the bonding modes, whereas the higher energy modes are the anti-bonding ones. These modes are collectively known as hybridized modes.

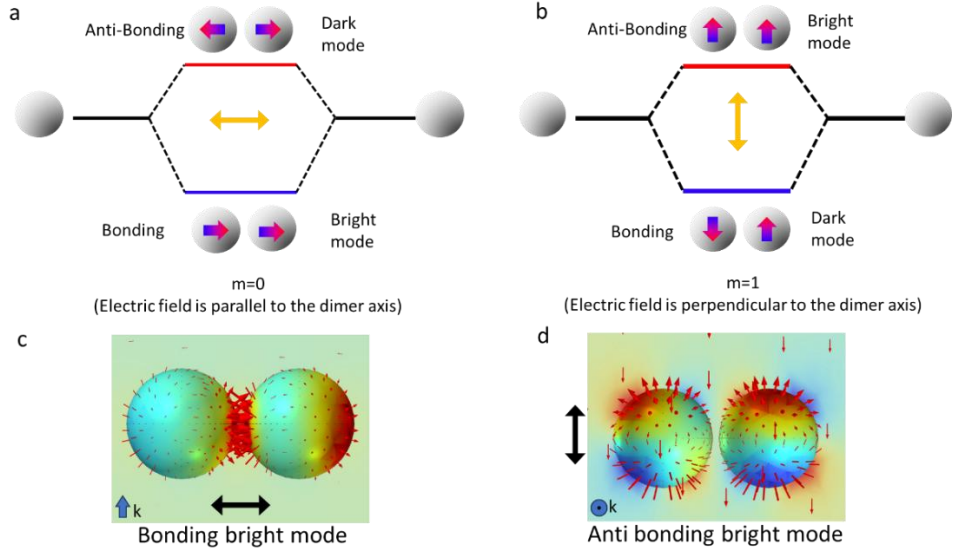

Fig. 3. The modes in nanoparticle dimer. a, The modes in a dimer when the electric field (polarisation depicted by the yellow doubled arrow) of the incident wave is parallel to the dimer axis. b, the modes in the dimer when the electric field (polarisation depicted by the yellow doubled arrow) is perpendicular to the dimer axis. The arrow in the spheres represents the direction of surface charge density from the negatively charged region to the positively charged region in the individual sphere. The blue and red regions represent negative and positive surface charge densities respectively. c, The surface charge density (relative) on Ga when the polarisation of the electric field (black double arrow) is in the direction of the dimer axis. The propagation direction of the incident plane wave is towards the top of the page. d, The surface charge density (relative) on Ga when the polarisation of the electric field (black double arrow) is perpendicular to the direction of the dimer axis. The propagation direction of the incident plane wave is out of the page.

The absorption efficiency for Ga dimers depends on the dimer gap. We define the gap as the shortest distance from the surface of the Ga nanosphere to the next. If $d$ is the distance between the centres of the dimers consisting of spheres with radii $r_1$ and $r_2$, then the gap is given by $g = d - r_1 - r_2$. If the sizes are comparable $r_1 \approx r_2 = r$, the gap is then $g = d - 2r$.

For the case of Ga dimers in PDMS, the absorption efficiency spectra have been calculated for both anti-bonding and bonding modes (see Figure 4a-b) as a function of the dimer gap. Figure 4c depicts the hybridized modes as a function of the dimer gap. The bonding mode energy lowers to the visible region from UV region as the gap decreases (Figure 4c). Equivalently, on increasing the gap, the bonding mode energy increases, thus leading to a blue-shift of the optical spectra.

The morphology of experimentally observed samples is equivalent to randomly oriented Ga dimers. Due to the large number of Ga nanodroplets and the incident light being unpolarised, the experimental spectra will have contributions from both the bright modes (m=0 and m=1). Henceforth, we shall refer to the bight modes only, and we shall refer to m equals 0 and m equals 1 modes as bonding and anti-bonding modes, respectively.

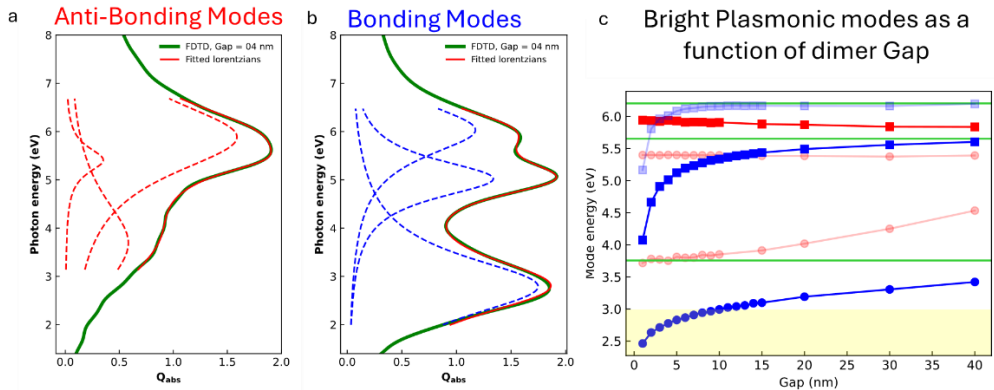

Fig. 4. Plasmonic Modes in Gallium Homo-Dimer Embedded in PDMS | The Absorption Efficiency of Ga Dimer with 4nm Gap in (a) Anti-Bonding Modes (the gap is perpendicular to the direction of the electric field of the incident plane wave), (b) Bonding Mode (gap is parallel to the direction of the electric field of the incident plane wave) (c) Peak of absorption efficiency as a function of gap.

We note that the experimentally observed trends of the dipole mode are similar to those expected from the Ga dimer simulations. With the increasing linear strain of the PDMS substrate, the inter-droplet gap between the Ga nanodroplets increased, leading to the blue shift of the experimentally observed reflectivity spectra as shown in Figure 5a. Furthermore, from Ga dimer simulation (SI Figure 9), the quality factor of dipole mode is expected to decrease with increasing inter-droplet gap. Experimentally we observe a decrease in quality factor (Figure 5b) from 2.94 (unstrained) to 2.56 at 45% strain. The dipole mode energy (Figure 5b) increases (blue-shifts) by a factor of 1.1 from 2.07 eV (unstrained) to 2.28 eV at 45% strain, while the full width at half maximum (FWHM) (SI Figure 10) increases by a factor of 1.25 from 0.71 eV (unstrained) to 0.89 eV resulting in the observed decrease of the quality factor of the dipole mode. Although the simulation of plasmonic modes shows the peaks of the absorption cross-section, the scattering cross-section also exhibits a blue shift with an increase in inter-droplet gap. The reflectivity spectra consist of the scattering spectra and are expected to behave similarly, as observed experimentally (Figure 5).

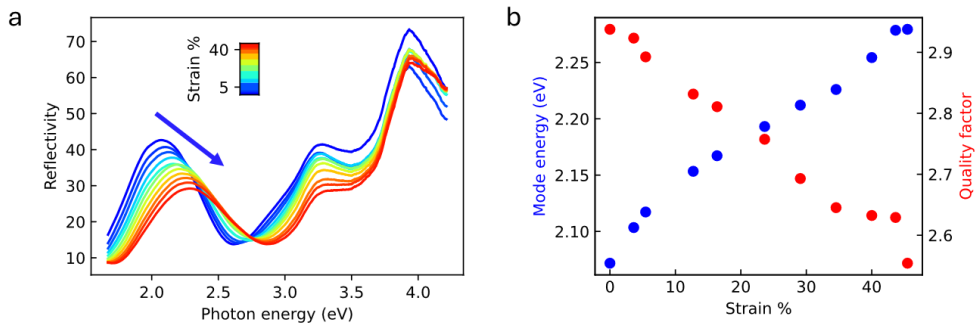

Fig. 5. Experimentally observed blue shift of reflectivity spectra with increasing strain. (a) Reflectivity spectra of the sample obtained experimentally. The arrow depicts the change of the dipole mode spectral peak with increasing strain. (b) Mode energy (shown in blue), and quality (shown in red) factor of dipole mode for different strains applied to the sample.

## 6. Conclusion

In conclusion, we have shown that non-coalescent Ga nanodroplets embedded in PDMS allow for hybridized plasmon modes. The bonding dipole modes are well within the visible region, enabling visual changes as the plasmon hybridization strength is altered by changing the inter-droplet gap. The tuneability is affected by a control change in the physical dimensions of the elastomer PDMS in which the Ga nanodroplets are embedded. The blue shift of reflectivity is the result of the reduced interaction strength of plasmons in neighboring electron-dense Ga droplets, which results in an increase of energy of the bright bonding mode. While the energetics of the plasmonic modes in nanoparticle dimers had been worked out as early as in 2004 [38], our sample demonstrates the real-time demonstrations of the effect of the same in liquid metals thus offering potential applications in soft photonics and robotics.

**Funding.** Prime Ministers Research Fellowship (TF/PMRF-21-1343), MHRD; SERB grant (SP/SERB-22-0021.05), DST India.

**Acknowledgment.** The authors acknowledge the Prime Minister Research Fellowship (TF/PMRF-21-1343), MHRD, SERB grant (SP/SERB-22-0021.05), and DST India for funding the project. They also thank the Micro and Nano Characterization Facility (MNCF), CENSE IISc, and Advanced Facility for Microscopy and Microanalysis (AFMM), IISc, Bangalore.

**Disclosure.** The authors declare no conflicts of interest.

**Data Availability.** Data underlying the results presented in this paper are not publicly available at this time but may be obtained from the authors upon reasonable request.

See Supplement 1 for supporting content.

# SUPPLEMENTARY INFORMATION

# Real-time tuneable bright bonding plasmonic modes in Ga nanostructures.
### RENU RAMAN SAHU[1], TAPAJYOTI DAS GUPTA[1]

*1Laboratory of Advanced Nanostructures for Photonics and Electronics, Department of Instrumentation and Applied Physics, Indian Institute of Science, Bangalore, India-560012*

## 1 Plasmonic Modes in Single Gallium Nanosphere

The plasmonic modes in a single Ga nanosphere of radius $r$ in PDMS medium (refractive index $n_{PDMS} = 1.4$) are obtained analytically from the Mie theory by calculating the absorption cross-section $\sigma_{abs}$ with units of area. Equivalently, the unitless quantity, absorption efficiency is also used, given by:

$$Q_{abs} = \frac{\sigma_{abs}}{\pi r^2}$$

The absorption efficiency ($Q_{abs}$), is the difference between the extinction efficiency ($Q_{ext}$) and the scattering efficiency ($Q_{sca}$)1:

$$Q_{abs} = Q_{ext} - Q_{sca}$$

From Mie theory,

$$Q_{ext} = \frac{2}{x^2} \sum_n (2n+1) \, \text{Re}\{a_n + b_n\}$$

$$Q_{sca} = \frac{2}{x^2} \sum_n (2n+1)(|a_n|^2 + |b_n|^2)$$

where $x = 2\pi r \, n_{PDMS}/\lambda$ is the size parameter when a plane wave of wavelength ($\lambda$) is incident on it. The Mie coefficients $a_n$ and $b_n$ represent the electric and magnetic multipoles indexed by the subscript $n$. For example, $a_1$, $b_1$, $a_2$ and $b_2$ represent the contribution from the electric dipole, magnetic dipole, electric quadrupole and magnetic quadrupole respectively.

Figure 1a depicts the absorption efficiency spectra for different radii of the Ga nanosphere. The entire spectra can be considered as a sum of the individual spectrum resulting from charge oscillation in a distribution akin to a spherical harmonic. The spectrum corresponding to each spherical harmonic mode is said to be contributed by that mode. The major contribution of the observed optical spectra comes from the electric dipole and quadrupole terms, with a small, non-negligible contribution coming from the electric octupole term (Figure 1b, 1c). The magnetic poles negligibly contribute to the spectra (Figure 1d). With the increase in radii, the spectrum redshifts with a decrease in the absorption efficiency and spectral broadening of the modes. Also, the octupole mode becomes more prominent in the spectra from higher radii Ga nanospheres. The radii of Ga nanospheres in our experiment are usually less than 50nm. Therefore, for the experimental and numerical spectra we obtain, the electric dipole and quadrupole terms suffice for analysis.

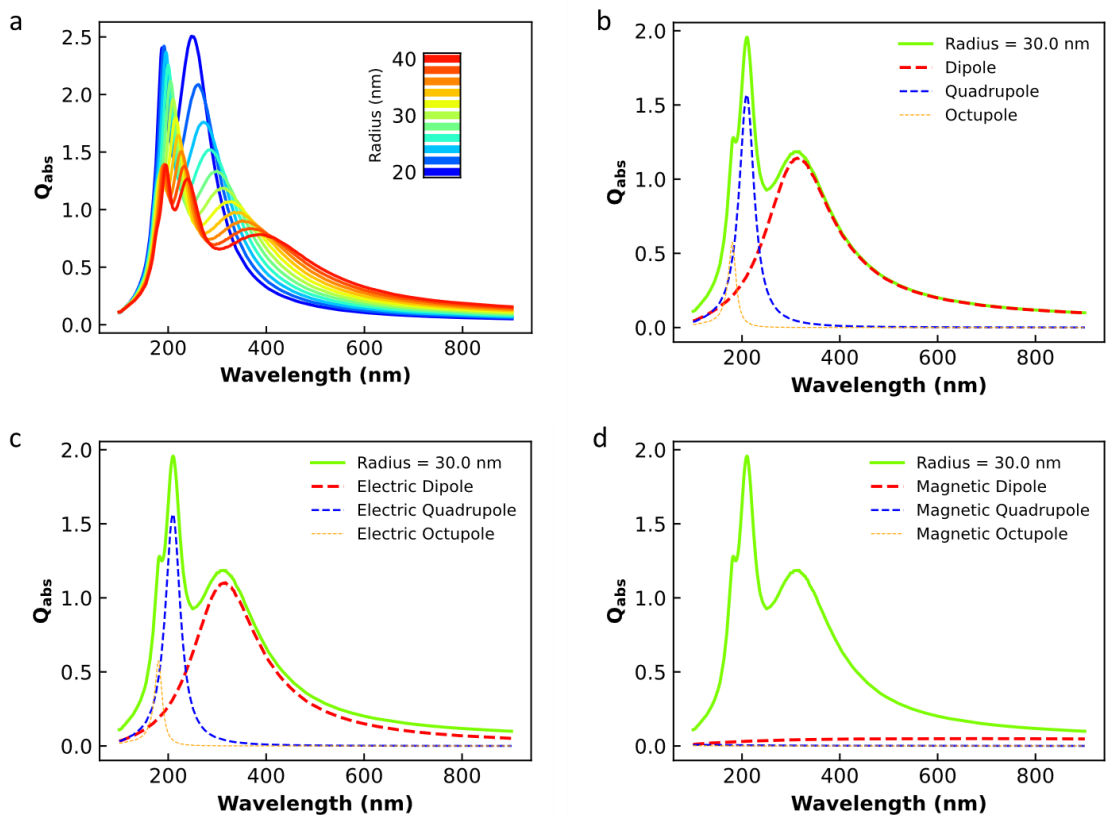

SI Figure 1: The plasmonic modes of a single Ga nanodroplet obtained from Mie theory. (a) Absorption scattering efficiency of Ga nanospheres with different radii. (b) Dipole, quadrupole, and octupole modes are present in the spectrum of absorption efficiency of Ga nanosphere of radius 30nm. (c) The contribution to the spectrum by electric dipole, electric quadrupole and electric octupole terms. (d) The contribution to the spectrum by magnetic dipole, magnetic quadrupole and magnetic octupole terms.

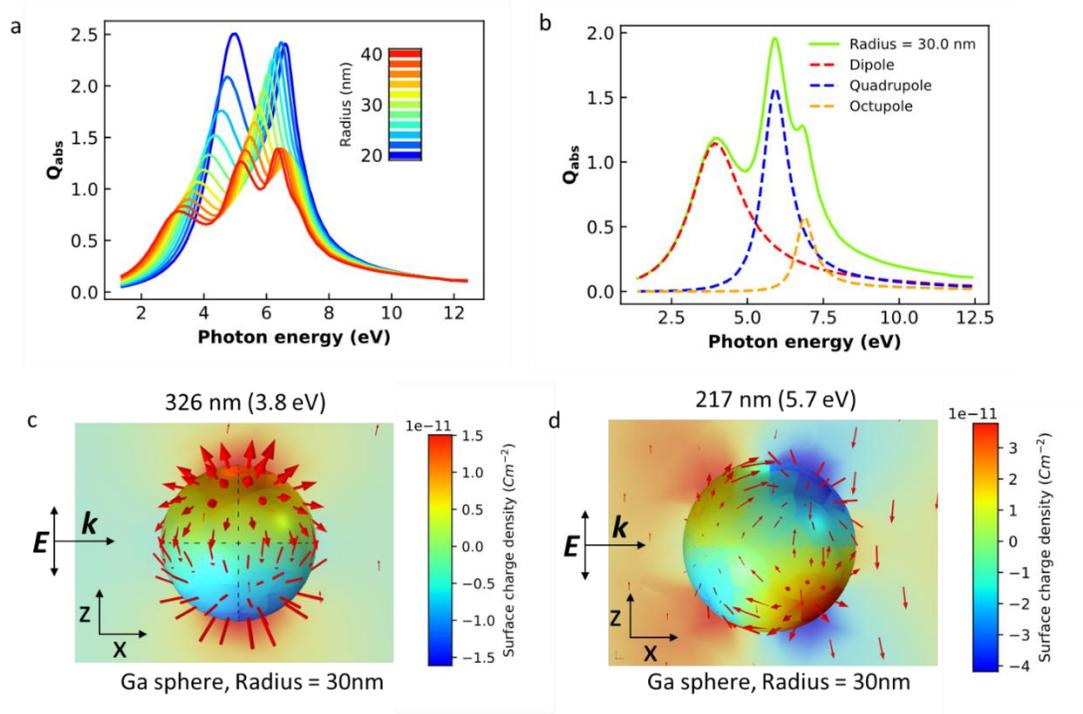

SI Figure 2: The plasmonic modes of a single Gallium nanodroplet obtained from Mie theory in the photon energy space. (a) Absorption scattering efficiency of Ga nanospheres with different radii. (b) Dipole, quadrupole, and octupole modes are present in the spectrum of absorption efficiency of Ga nanosphere of radius 30nm. (c) The surface charge density on the Ga sphere placed in PDMS medium (shown by the colour legend on the sphere) when a plane wave of vacuum wavelength 326 nm (3.8 eV photon energy) is incident on it. The incident plane wave is polarised linearly in the z-direction and propagates along the x-axis. The arrows represent the direction of the electric field. (d) The surface charge density on the Ga sphere placed in PDMS medium (shown by the colour legend on the sphere) when a plane wave of vacuum wavelength 217 nm (5.7 eV photon energy) is incident on it.

The spectral shape in the photon energy space can be approximated with the Lorentzian line shape corresponding to each mode. Figure 2a depicts the absorption efficiency spectra for different radii of the Ga nanosphere in the photon energy space. Note that the modes of the spectra, as calculated from Mie theory, are symmetric and have a Lorentzian line shape (Figure 2b). The center of these Lorentzians is the energy of the mode of the single sphere.

The surface charge distribution at the peak of dipole mode and that at quadrupole mode have a surface charge density, which when decomposed into spherical harmonics, has a major contribution from the component $Y(l=1, m=0)$ and $Y(l=2, m=\pm 1)$ respectively as shown in Figure 2c, 2d and Figure 3. The $m$ component is determined by the geometry of the plane wave illumination. The coefficient of a spherical harmonic $Y_{lm}(\theta, \phi)$ mode present in the surface charge density $\sigma_s(\theta, \phi)$ is given by

$$c_{lm} = \int_{\theta=0}^{\pi} \int_{\phi=0}^{2\pi} \sigma_s(\theta, \phi) Y_{lm}(\theta, \phi) \, d\theta d\phi$$

In general, $c_{lm}$ is a complex number. The relative contribution (in %) of the mode is calculated by

$$c_{lm}^{nomalized} = \frac{|c_{lm}|}{\sum_l \sum_{m=-l}^{+l}|c_{lm}|} \times 100$$

When 30nm Ga nanosphere in PDMS medium is illuminated with plane wave of photon energy 3.8eV, the surface charge density is dominated by the dipole contribution (Figure 3a) with almost 75%. The rest of the contribution comes from the quadrupole terms. When illuminated with a photon energy of 5.7eV photon energy, the quadrupole contribution is about 74% and the rest is from dipole terms.

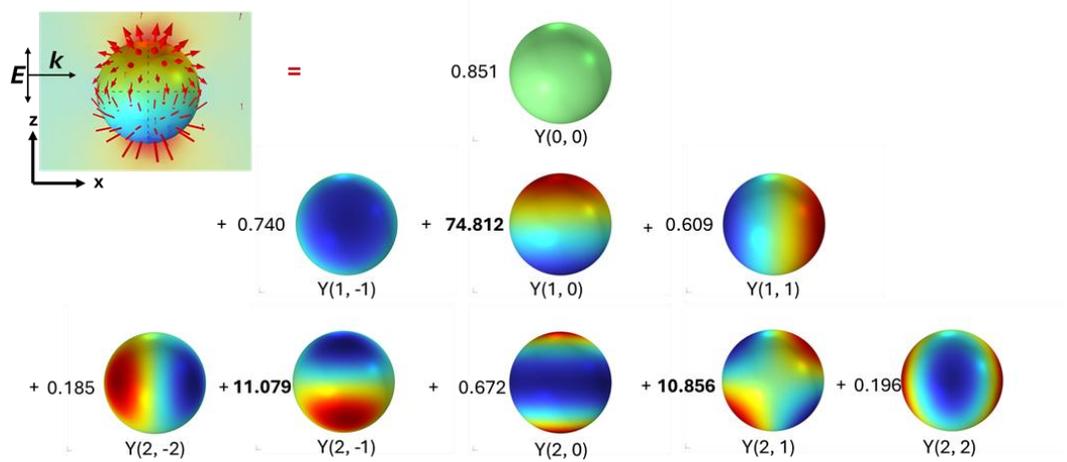

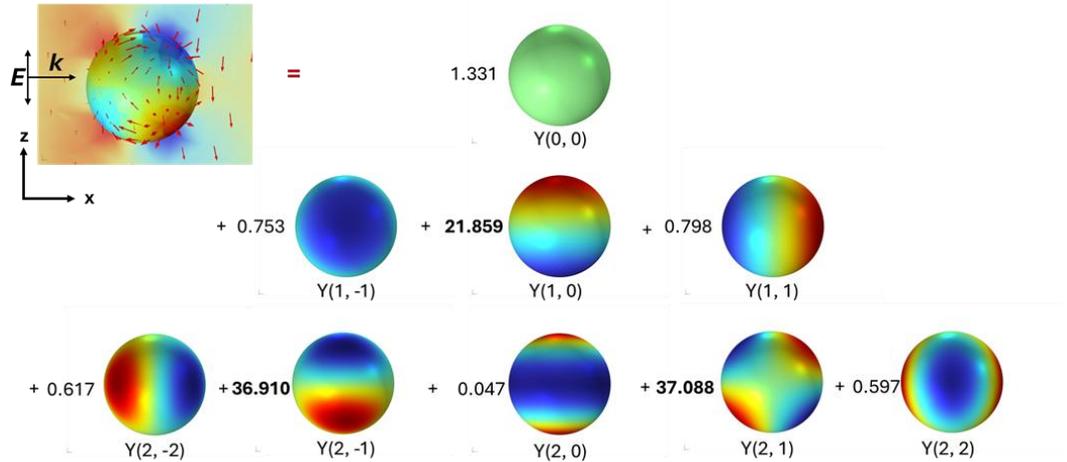

SI Figure 3: Spherical harmonic decomposition of surface charge density of 30nm radius Ga nanosphere at the (a) dipole peak (326nm), and (b) quadrupole peak (217 nm). The relative magnitude of the coefficient is against each spherical harmonic Y(l, m). The normalisation is chosen by taking the sum of the coefficients to be 100. The higher order modes are negligibly small and hence neglected.

## 2 Determination of the Mode of a spectral peak

To determine the mode corresponding to a peak in the spectrum, we fit the peak with a Lorentzian and obtain the central photon energy. The surface charge density is determined corresponding to the plane wave illumination at that photon energy. The spherical harmonic contribution is determined using the above formula for $c_{lm}^{nomalized}$. The mode corresponding to the peak corresponding to the largest value of $c_{lm}^{nomalized}$ (Figure 4).

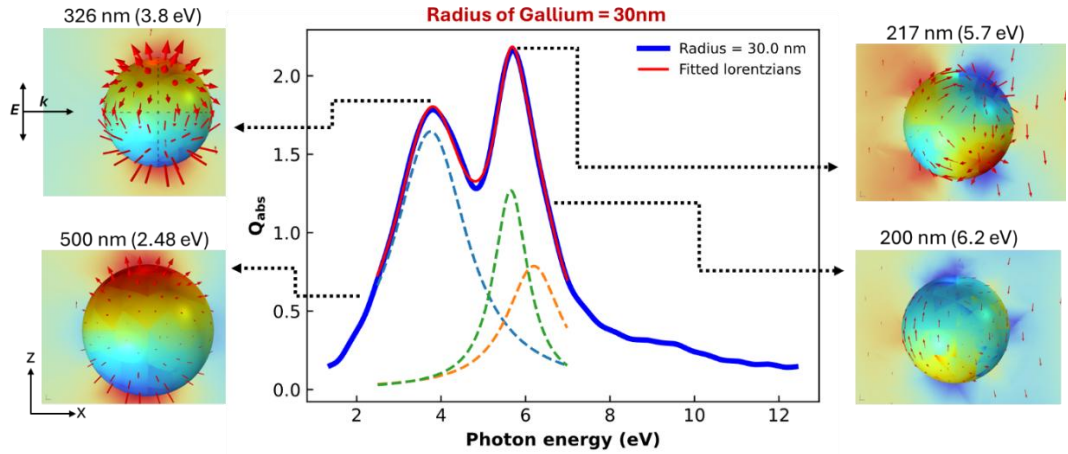

SI Figure 4: Absorption efficiency spectra of a single Ga nanosphere of radius 30nm from FDTD simulation. The individual modes are obtained by decomposition of the spectra into Lorentzians. The surface charge density of the Ga nanosphere on plane wave illumination at photon energies 2.48eV, 3.8eV, 5.7eV and 6.2eV are shown, with the dotted arrows representing the corresponding absorption efficiency at that photon energy. Note that at the lowest energy local maxima of the spectrum, the dipole moment is the largest dipole moment, as indicated by the relative sizes of the arrows representing the electric field. The higher energy peak corresponds to the quadrupole mode, while the octupole mode peak is quite small (hidden due to numerical errors in the simulation but can be observed as a subtle peak from Mie theory calculations as shown in the previous figure).

### 3 Frolich's condition for Ga nanodroplets

In the subwavelength size limit, the polarizability of a nanosphere is maximized at Frolich's condition2

$$\epsilon_{Ga} = -2\epsilon_m$$

where $\epsilon_{Ga}$ is the permittivity of Ga and $\epsilon_m$ is the permittivity of the surrounding medium in which Ga nanodroplets are embedded. The dominant contribution to the optical spectra comes from the wavelength of light (photon energy) at which the polarizability is the maximum and gives the energy of dipole surface plasmon mode.

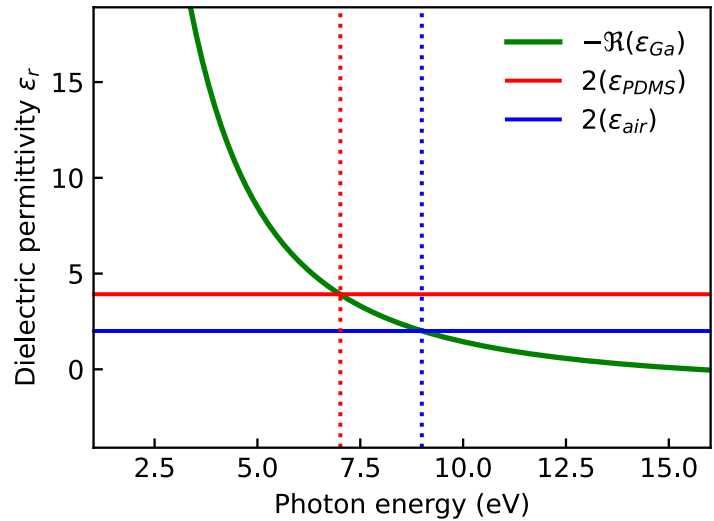

SI Figure 5: Frolich's condition for Ga embedded in air and PDMS. When embedded in a PDMS medium the LSPR mode of single Ga nanodroplet is expected to be observed at 8.9eV. With an increase in the refractive index of the surrounding medium, the LSPR photon energy decreases.

In the air medium, the LSPR occurs at 8.9 eV (Figure 7). As the refractive index of the medium increases, the LSP energy decreases towards 7eV. According to Mie theory, it is well known that with a decrease in the particle size, the resonant energy increases (blue-shifts). The upper bound to this blue shift is set by Frolich's condition. Arbitrarily decreasing the particles lead to an increase of the localized dipole resonance only up to the energy given by Frolich's condition.

## 4 Size dependence of optical cross-sections

Ga nanodroplets much smaller than subwavelength sizes experience spatial homogeneity as compared to the larger ones They contain only dipole mode as a major mode of plasmon oscillation. Higher-order modes, such as quadrupole and octupole modes, become significant as the size of the scatterer becomes comparable to that of the visible light (SI Figure 6:). Ga spheres of 20nm radius have two peaks. The lower energy peak corresponds to the dipole mode and has a larger amplitude as compared to the peak at higher energy which corresponds to the quadrupole mode. As the size of the Ga sphere increases to a 40nm radius, an additional peak comes up, indicating a significant contribution from the octupole mode.

The size of the nanodroplet, in our case, is usually sub-100nm, typically in the radii ranging between 20nm to 40nm. As shown in the extinction, scattering, and absorption spectra, the contribution from the dipole mode is dominant, and that from the octupole mode is minimal. Therefore, in the study of the dimer modes, we fit the spectra with those many numbers of lorentzians as the number of visible peaks and label them as contributions from the dipole, quadrupole, and octupole modes in order from the lower to higher photon energy.

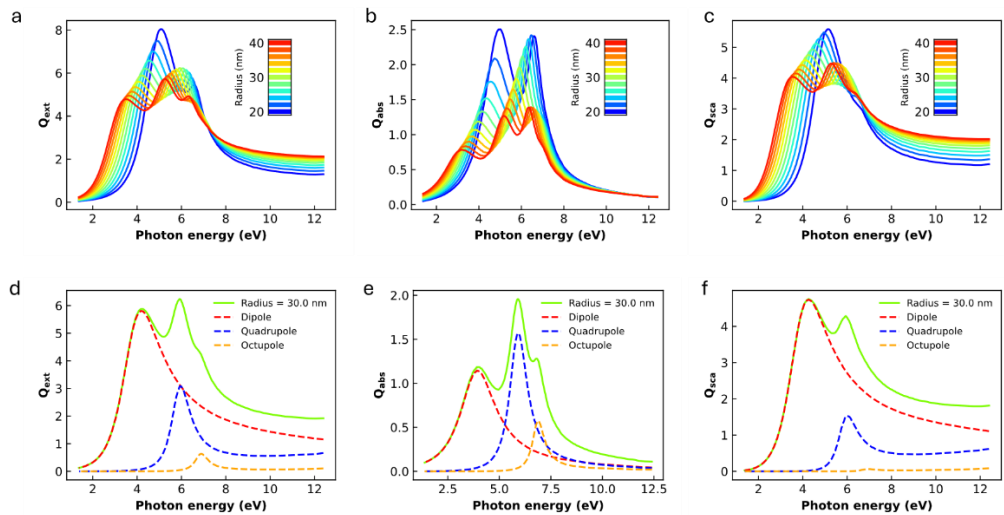

SI Figure 6: Optical cross-sections dependence on the size of Ga nanodroplet. (a) Extinction, (b) Absorption, and (c) Scattering cross-sections. (d), (e), and (f) depict the decomposition of the extinction, absorption and scattering cross sections respectively into the contributions from dipole, quadrupole and octupole modes.

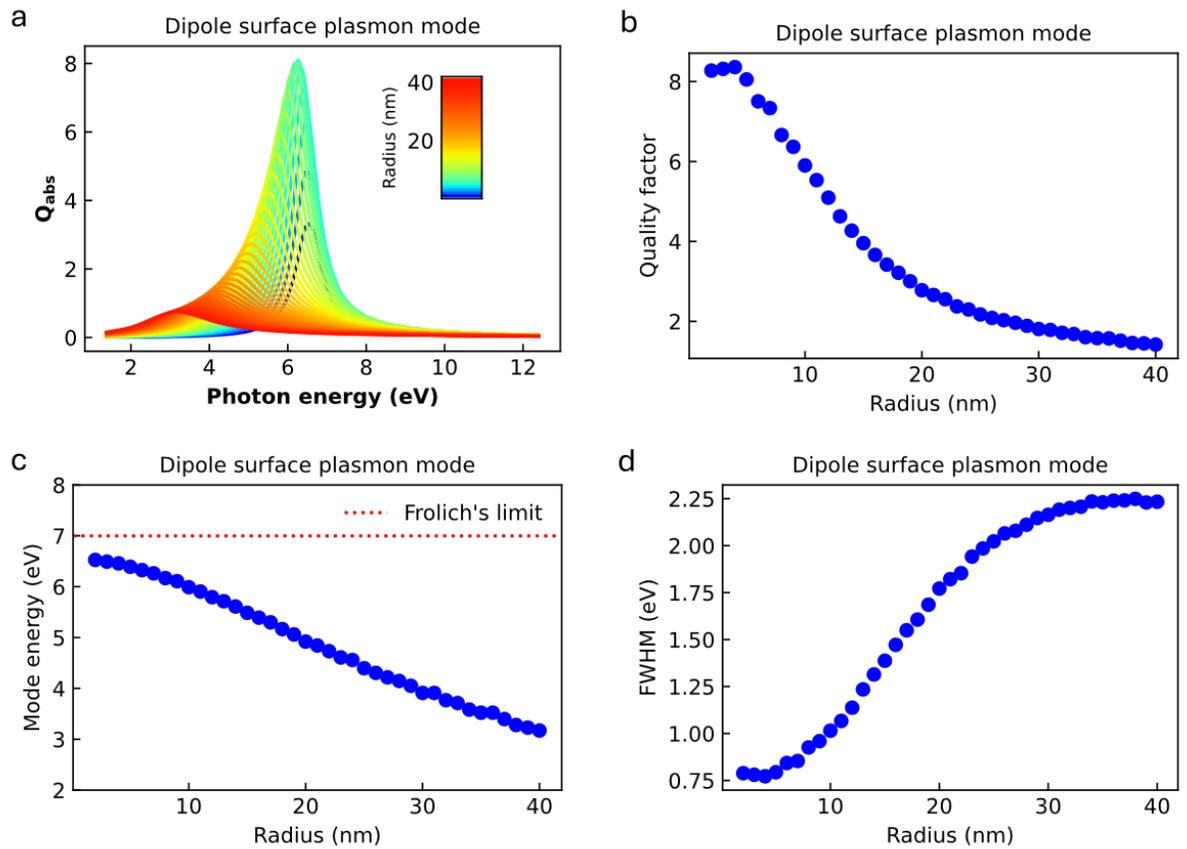

SI Figure 7: Dipole surface plasmon mode of single Ga nanodroplet. (a) Size dependence of the dipole surface plasmon mode of single Ga nanodroplet. The (b) quality factor, (c) Mode energy, (d) Full width at half maxima (FWHM) of dipole surface plasmon mode.

## 5 The blue shift of optical cross-sections with an increase in inter droplet gap

The trend in a spectral shift as a function of the inter-droplet gap is the same for extinction, absorption, and scattering cross-sections. In particular, with an increase in inter-droplet gap, all the aforementioned spectra blue-shift (SI Figure 6).

In a dimer configuration, the plasmonic peaks denoted by the peak of absorption cross-section shift towards higher energy owing to an increase in energy of bonding dipole mode. The plasmon modes are fixed for a given geometrical configuration. Therefore, all the optical spectra, including the absorption, scattering, and extinction cross-section, exhibit a shift towards higher energy as the inter-droplet gap increases, owing to a decrease in the plasmon hybridization strength.

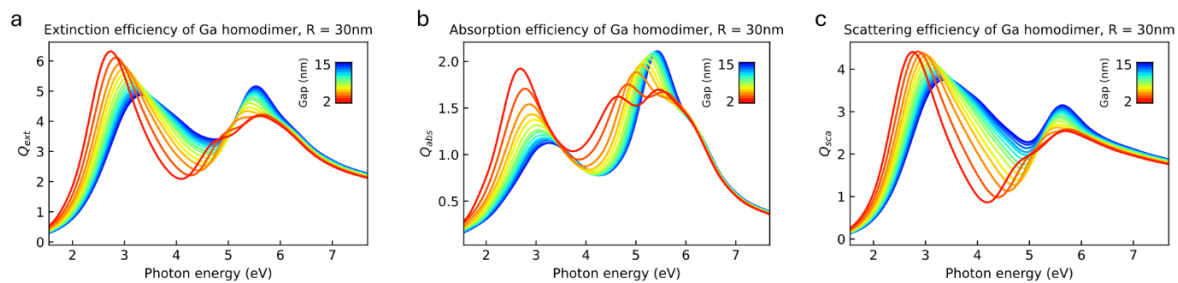

SI Figure 8: Shift of optical spectra (a) Extinction efficiency, (b) Absorption efficiency, (c) Scattering efficiency, from Ga homodimers with spheres of radius 30nm as the inter-droplet gap increases.

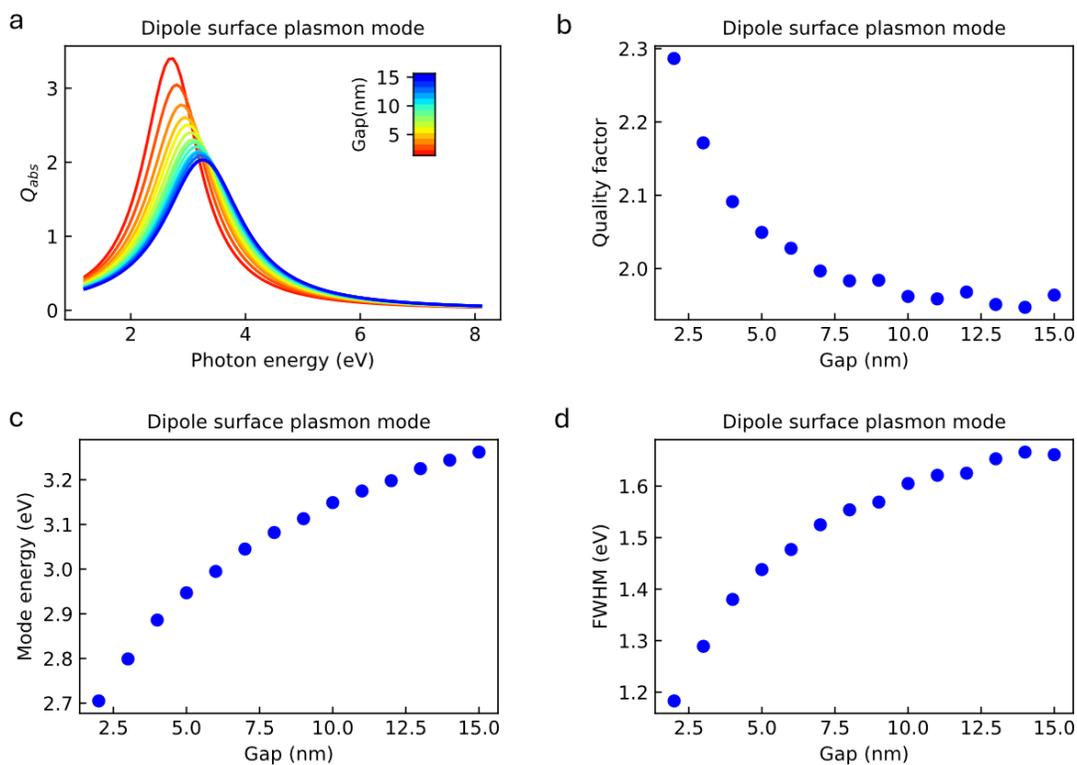

SI Figure 9: Bright-bonding mode of Ga dimer. (a) Gap dependence of the mode of single Ga nanodroplet. The (b) quality factor, (c) Mode energy, (d) Full width at half maxima (FWHM) of the mode.

## 6 Quality factor of dipole mode from experimental spectra

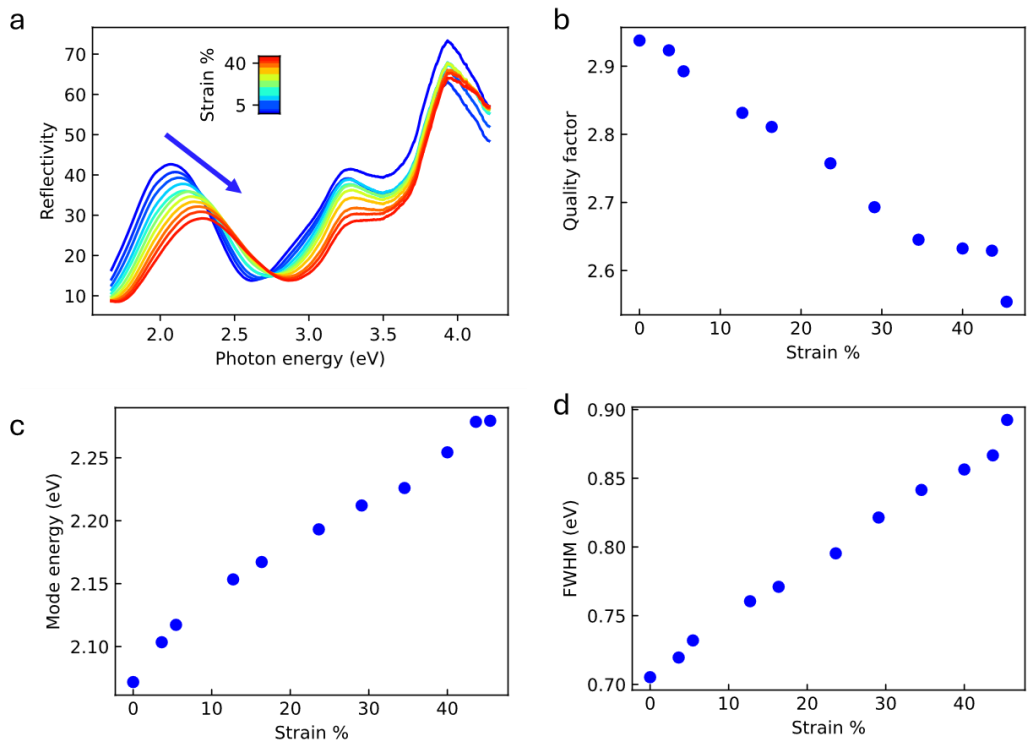

SI Figure 10: Experimentally observed blue shift of reflectivity spectra with increasing strain. (a) Reflectivity spectra of the sample obtained experimentally. The arrow depicts the change of the dipole mode spectral peak with increasing strain. (b) Quality factor, (c) mode energy and (d) FWHM of dipole mode for different strains applied to the sample.